# A Compact Size 5G Hairpin Bandpass Filter with Multilayer Coupled Line

**Qazwan Abdullah [*,1,2], Ömer Aydoğdu[2], Adeeb Salh[3], Nabil Farah[4], Md Hairul Nizam Talib[4], Taha Sadeq[5], Mohammed A. A. Al-Mekhalfi, Abdu Saif [6]**

[1]Faculty of Electrical and Electronic Engineering, Selçuk University, Konya, Turkey
[2]Faculty of Engineering and Natural Sciences, Konya Technical University, Konya, Turkey
[3]Faculty of Engineering Technology, Universiti Tun Hussein Onn Malaysia, Pagoh, Muar, Johor, Malaysia
[4]Faculty of Electrical Engineering, Universiti Teknikal Malaysia Melaka, Melaka, Malaysia
[5]Faculty of Engineering and Science, Universiti Tunku Abdul Rahman, Sungai Long Campus, Selangor 43000, Malaysia
[6]Faculty of Electrical Engineering Department, University of Malaya, Kuala Lumpur, Malaysia
[*]Corresponding Author: Qazwan Abdullah. Email: gazwan20062015@gmail.com
Received: XX Month 202X; Accepted: XX Month 202X

**Abstract:** The multilayer structure is a promising technique used to minimize the size of planar microstrip filters. In the flexible design and incorporation of other microwave components, multilayer band-pass filter results in better and enhanced dimensions. This paper introduces a microstrip fifth-generation (5G) low-frequency band of 2.52-2.65 GHz using a parallel-coupled line (PCL) Bandpass filter and multilayer (ML) hairpin bandpass filter. The targeted four-pole resonator has a center frequency of 2.58 GHz with a bandwidth of 130 MHz. The filters are designed with a 0.1 dB passband ripple with a Chebyshev response. The hairpin-line offers compact filter design structures. Theoretically, they can be obtained by bending half-wavelength resonator resonators with parallel couplings into a "U" shape. The proposed configuration of the parallel-coupled line resonator is used to design the ML band-pass filter. The FR4 substrate with a dielectric constant ($\varepsilon_r$) of 4.3 and 1.6 mm thickness was used. A comparative analysis between the simulated insertion loss and the reflection coefficient of substrates RO3003 and FR4 was performed to validate the efficiency of the proposed filter design. Simulation of PCL filter is accomplished using computer simulation technology (CST) and an advanced design system (ADS) software. The PCL Bandpass filter was experimentally validated and a total tally between simulation results and measured results were achieved demonstrating a well-measured reflection coefficient. The simulated results obtained by the hairpin ML bandpass filter show that the circuit performs well in terms of Scattering(S) parameters and the filter size is significantly reduced.

## 1 Introduction

There are growing future demands to further boost the performance of microwave technology. Various applications implement microwave communication including satellite broadcasting radar signals, phones, and navigational applications, millimeter-wave applications, and so on [1-3]. Microwave communication is the transmission of signals or powers from one point to another through microwaves. It allows the transmission of a vast amount of data among remote communication points at the same frequencies [4]. The electromagnetic spectrum's microwave range is limited and needs to be shared. One of the main components of a microwave communication system is filters which are two-port networks applied to control radio frequency (RF) response systems by allowing passband frequency transmissions and attenuating signals beyond their bandwidth [5]. A microstrip filter is then used to choose or enclose radio frequency (RF) or microwave signals within the spectrum limit allocated. Various types of filters can be used in a microwave communication system including low-pass, band-pass, band-stop, and high-pass filters. Among these types of filters, the band-pass filter is preferred in microwave systems due to its small size which makes it convenient for different applications, signal to noise ratio (SNR), and receiver sensitivity is also improved with the aid of the bandpass filtering method and gives stability and reliability [6]. There are two types of filters which they are active and passive. In this research, a passive filter serves as an attenuator, thereby supplying an output signal with a lower amplitude compared to the input signal. However, certain amplification steps must be considered when designing the circuitry to provide a distortionless signal at the output. The ongoing advancements in wireless communication systems require effective and high-performance band-pass filters. Parallel-coupled line (PCL) microstrip filters are a type of band-pass filter that is most commonly used in various wireless communication and microwave systems [7-8]. PCL microstrip filters have the features of a planar structure, less fabrication complexity, low cost, and relatively wide bandwidth [9]. Various parallel-coupled filter structures have been reported in the literature [10-11]. However, developing a high-performance PCL band-pass filter with a compact size has been attracting a wide concern of researchers. The S-parameter performance like return loss ($S_{11}$) and insertion loss ($S_{21}$) considered a critical issue of PCL band-pass filters [12-13]. To realize the significance of compact size, less reflection, and insertion losses of PCL band-pass filter, the digital broadcasting application is considered.

Digital broadcasting denotes a set of standards aimed at the precise and structured delivery of transmitted signals in digital form [14]. Many band-pass filters applied in digital broadcasting do not use external power sources and consist only of passive components, such as condensers and inductors called passive band-pass filters. An effective band-pass filter has low insertion losses and offers a high degree of passband rejection by integrating multiple transmission zeros. However, the development of high-performance, compact size, lightweight, and low-cost band-pass microwave RF filters for evolving digital broadcasting remains a challenging issue in the field of communication systems[15]. Thus, a multilayer (ML) structure approach is introduced as an effective method in reducing the size of microwave devices since most communications systems end up with a portable consumer convenience device [16-17]. With the implementation of the multilayer structure and using high dielectric constant substrates, miniaturized filters can be produced [18-22]. There various studies that have considered the multilayer approach to achieve a compact size filter. A sample and low-cost ML filter design have been introduced in [23]. However, the PCL design and the filter parameters calculation were not presented. Also, [24] has proposed the ML filter design with different structures and substrates and has managed to reduce the filter size, but using different substrates may increase the system complexity. ML hairpin bandpass filter for digital broadcasting in [25]. The authors claimed that hairpin shape is based on bending the resonator of PCL which was not presented. The calculation of physical filter parameters like space (S), width (W), and length (L) was not shown. The size reduction was not presented as well. A compact ML bandpass filter with a modified hairpin resonator in [26] was presented well. However, the size reduction was not enough. He claimed that the size of the filter has reduced 50% from the single layer. However, the size is still not small compared to [27] and [28] in the same frequency range, and bandwidth is limited due to gap limitation.

This research concentrates on developing a band-pass filter with four poles hairpin using the ML configuration to reduce the size of microstrip filters. The arranging of multiple layers of microstrip lines on the same substrates and overlapping these lines can achieve a strong coupling filter [29]. The adjacent hairpin resonator lines are placed at different levels to change coupling strength by varying the overlapping gap between two resonators vertically, this is beneficial for reducing the filter length and obtaining asymmetrical response [30]. Besides, with a multilayer configuration, any shape of the resonator for each layer can be selected, then these layers are combined to produce the ML filter structure [31]. Besides, the hairpin resonator was combined in the ML configuration with half-wavelength coupled line resonators. The resonator design is mounted on two same substrates with the same dielectric constants, placing the hairpin resonators on the bottom layer and the top layer. The hairpin structure was chosen due to its advantages in reducing the size and minimizing the cost of the filter design [29-31]. The study also provides a comprehensive discussion of the calculation of filter parameters and PCL design based on different materials. To validate the effectiveness and show the reduced size of the hairpin ML band-pass filter, the designs are carried out using CST software. This study will be a part of our fifth-generation (5G) ultra-wideband multiple input multiple outputs (UWB-MIMO) antennas in the dissipative media project. The paper is organized as follows: Section II material and methods, PCL Bandpass Filter Design, Agilent ADS simulation, CST-MW studio simulation and measurement, ML bandpass filter topology, and the physical layout of ML hairpin bandpass filter: Section III, FR4 based design, Rogers RO3003 based design and parametric studies and Section IV summarizes the study and highlights the findings and outcomes of the study.

**2 System Design**

The procedure begins with the filter specification and the type of ripple level expected. A microstrip bandpass filter has been designed with fractional bandwidth (FBW) = 0.05038 at a mid-band frequency of $f_0$= 2.58 (GHz). The design starts with a filter order calculation referring to Eq.1, Eq. (2), Eq. (3), Eq. (4), and Eq. (5). Four poles (n = 4) are the filter order and Chebyshev low-pass prototype with a 0.01dB passband ripple has been selected. The design begins with the low-pass filter consisting of series and parallel sub-inverter used to convert low-pass to band-pass filter prototype using the shunt branch. The parameters of the low-pass prototype for normalized low-pass cut-off frequency is $\Omega c=1$, where $g_0=1$, $g_1= 0.7129$, $g_2=1.2004$, $g_3=1.3213$, $g_4= 0.6476$ and $g_5= 1.1007$.

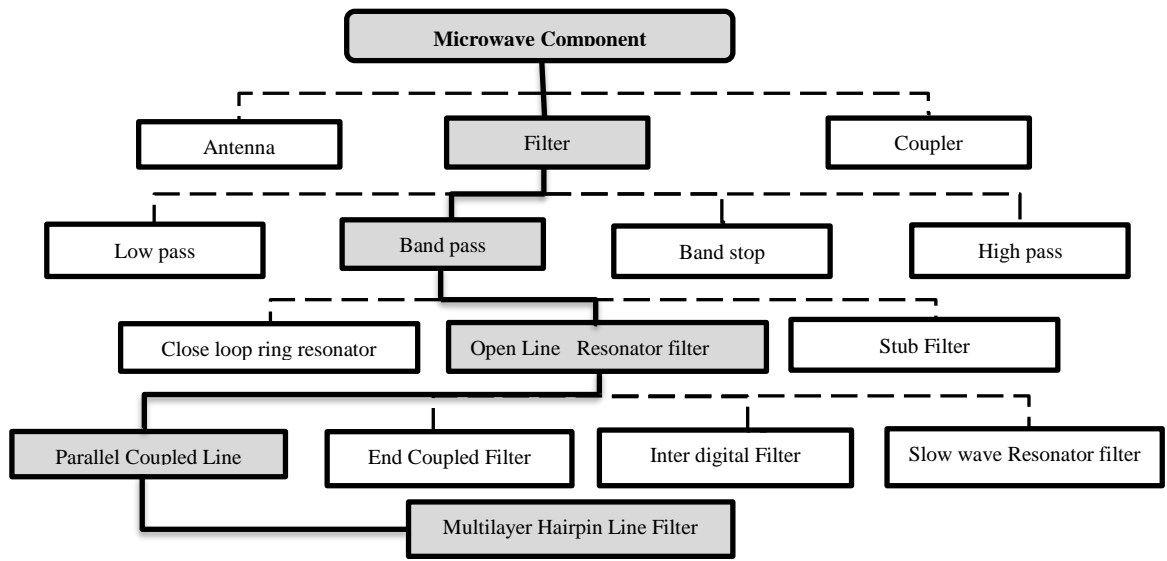

**Figure 1:** Summary of research flow.

The next step in the filter design is to find the dimensions of coupling microstrip lines that display the desired even and odd mode impedances. The important procedure is to find the physical dimension that can be divided into the coupling section, the admission inverter, $z_{0e}$ is even $z_{0o}$ impedance, space, width, and length. Fig. 1 illustrates the flow of the research where the bold lines reflect the system or configuration chosen to achieve the goals and the dotted lines reflect certain devices or configurations not addressed

For ripple 0.01

$$0.01\text{dB} = 10\log(a_m^2 + 1) \qquad (1)$$

where the ripple height,

$$a_m^2 = 10^{0.001} - 1 \qquad (2)$$

where attenuation height,

$$a = \sqrt{\frac{(10^{2.5}-1)}{a_m^2}} \tag{3}$$

Low-frequency transformation to band-pass, where $\omega_0$ is the center frequency and $\omega xl$ is the normalized frequency, where $\omega_1$ is lower frequency range, $\omega_2$ is the upper-frequency range and $\omega_x$ is the attenuation response.

$$\Omega xl = \frac{\omega_0}{\omega_2 - \omega_1}\left(\frac{\omega_x}{\omega_0} - \frac{\omega_0}{\omega_x}\right) \tag{4}$$

$$\omega xl = \frac{2.58}{2.52 - 2.56}\left(\frac{2.77}{2.58} - \frac{2.58}{2.77}\right)$$

Filter order

$$n = \frac{\operatorname{acosh}(a)}{\operatorname{acos}(\omega xl)} = 4 \tag{5}$$

For the first coupling section

$$\frac{J_{01}}{Y_0} = \sqrt{\frac{\pi}{2}\frac{FBW}{g_0 g_1}} \tag{6}$$

For the intermediate coupling section

$$\frac{J_{k+1,k}}{Y_0} = \frac{\pi FBW}{2}\frac{1}{\sqrt{g_k g_{k+1}}} \quad k = 1 \text{ to } n-1 \tag{7}$$

For the final coupling section

$$\frac{J_{n,n+1}}{Y_0} = \sqrt{\frac{\pi}{2}\frac{FBW}{g_n g_{n+1}}} \tag{8}$$

where $g_0, g_1 \ldots \ldots \ldots g_n$ are the elements of a ladder-type low-pass prototype with normalized cut-off $\Omega c = 1$, and FBW is the fractional bandwidth of the band-pass filter. J, j+1 is the characteristic admittances of J-inverters and $Y_0$ is the characteristic admittance of the lines [11]. Eq. (6), Eq. (7), and Eq. (8) are used to calculate the J-inverters. The even and odd line pair impedance was calculated using Eq. 9 and Eq.10 using the J admittance.

$$(z_{0e})_{j,j+1} = \frac{1}{Y_0}\left[1 + \frac{J_{j,j+1}}{Y_0} + \left(\frac{J_{j,j+1}}{Y_0}\right)^2\right] \quad j = 0 \text{ to } n \tag{9}$$

$$(z_{0o})_{j,j+1} = \frac{1}{Y_0}\left[1 - \frac{J_{j,j+1}}{Y_0} + \left(\frac{J_{j,j+1}}{Y_0}\right)^2\right] \quad j = 0 \text{ to } n \tag{10}$$

where $z_{0e}$ is even characteristic impedance and $z_{0o}$ is odd characteristic impedance. $Z_o = \frac{1}{Y}$ is the input and output line characteristic. Tab.1 is the summary of calculations for inverter admittances, even, and odd mode impedances. The line dimension in the microstrip filter is determined from the knowledge of even and odd mode coupling line admittance given in terms of the inverter impedance.

**Table 1:** Design parameters

| J | $J_{i+1/y_o}$ | $(z_{oe}), j_{i+1(n)}$ | $(z_{oo}) j, j_{i+1}$ |
|---|---|---|---|
| 0 | 0.3332 | 72.2092 | 38.8915 |
| 1 | 0.0855 | 54.6432 | 46.0886 |
| 2 | 0.0628 | 53.3393 | 47.0556 |
| 3 | 0.0856 | 54.6435 | 46.0884 |
| 4 | 0.3332 | 72.2092 | 38.8915 |

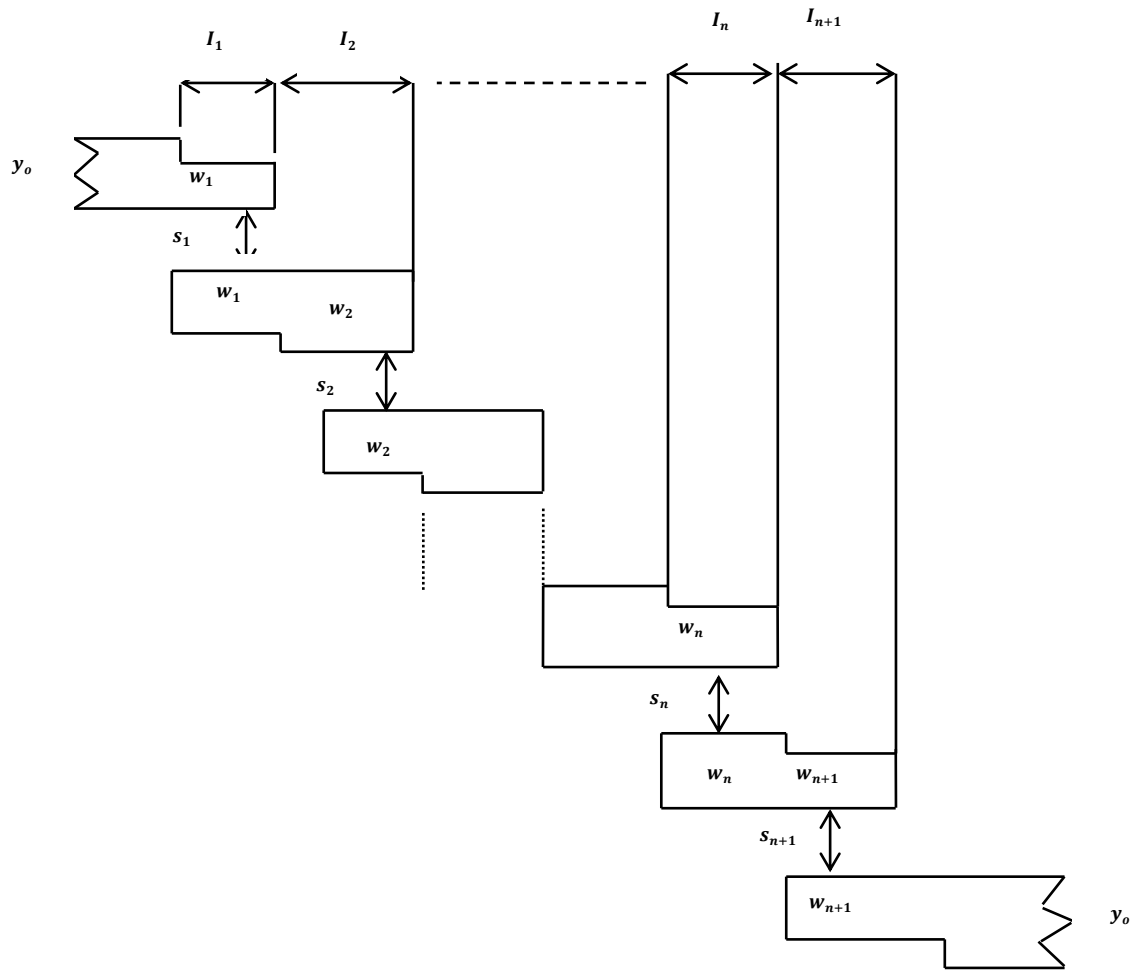

**Figure 2:** The general structure of parallel (edge)-coupled microstrip band-pass filter.

## 2.1 PCL Bandpass Filter Design

A PCL microstrip band-pass filter is also known as an edge-coupled band-pass filter. A typical PCL band-pass filter is made up of a bunch of parallel line resonators along their length with appropriate spacing between them. Hence, this creates a good microwave signal coupling between the neighboring resonators. In this study, a fourth-order band-pass filter, which is based on the Chebyshev filter response was carried out. A Chebyshev filter response was chosen because of its selectivity compared to Butterworth's response. The general structure of the PCL microstrip band-pass filter has been illustrated in Fig.2 which the parallel-coupled line width (W), length (L), gap (S), and impedance ($Y_0$) has been labeled accordingly.

### 2.1.1 Agilent ADS Simulation

The calculation of resonator dimensions can be done with the help of the ADS (Advanced Design System) using the special "Line Calc,''. The odd and even mode values in Tab. 1 are required to perform the calculation as shown in Figs. 3 and 4 for both materials FR4 and RO3003. The calculated dimensions from ADS have been summarized in Tabs. 2 and 3.

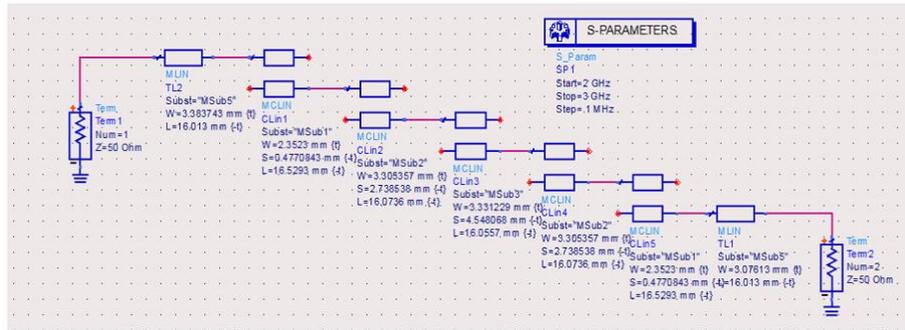

**Figure 3:** Parallel coupled line band-pass filter model (FR4).

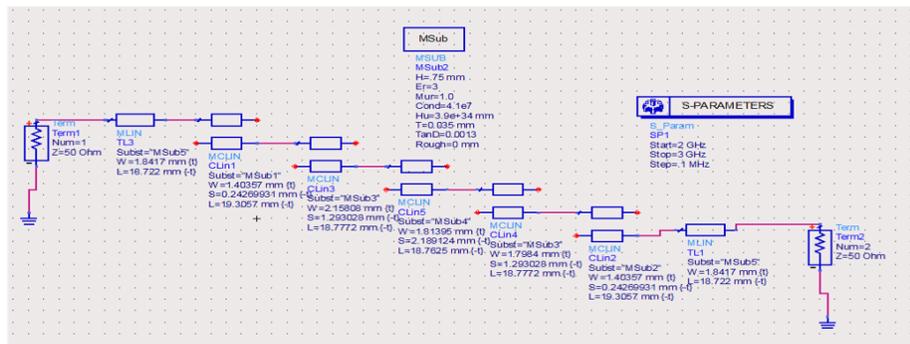

**Figure 4:** Parallel coupled line band-pass filter model (RO3003).

Table 2: Parameters of PCL filter using FR4

| n | W(mm) | L(mm) | S(mm) |
|---|---|---|---|
| 0 | 2.352 | 16.528 | 0.47708 |
| 1 | 3.305 | 16.073 | 2.738 |
| 2 | 3.331 | 16.0557 | 4.54 |
| 3 | 3.305 | 16.073 | 2.738 |
| 4 | 2.352 | 16.528 | 0.47708 |

Table 3: Parameters of PCL filter using (RO3003)

| n | W(mm) | L(mm) | S(mm) |
|---|---|---|---|
| 0 | 1.543 | 19.3057 | 0.28032 |
| 1 | 1.798 | 18.7772 | 1.2930 |
| 2 | 1.995 | 18.7625 | 2.0327 |
| 3 | 1.798 | 18.7772 | 1.2930 |
| 4 | 1.543 | 19.3057 | 0.28032 |

*2.1.2 CST-MW Studio Simulation and Measurement*

The PCL filter has been designed and simulated with the aid of dimensions of Tab. 2 and 3. The filter dimensions have been optimized to achieve the required response, which meets the design specifications. Figs. 5 and 6 display the filter's physical structure after optimization using R03003 and FR4 respectively. The total filter size is 128.26 mm x 21.57 mm and 73.765 mm x 27.506 mm respectively. Fig .7 demonstrates the rendered PCL filter. The filter's scattering parameters are measured via a Vector Network Analyser (VNA).

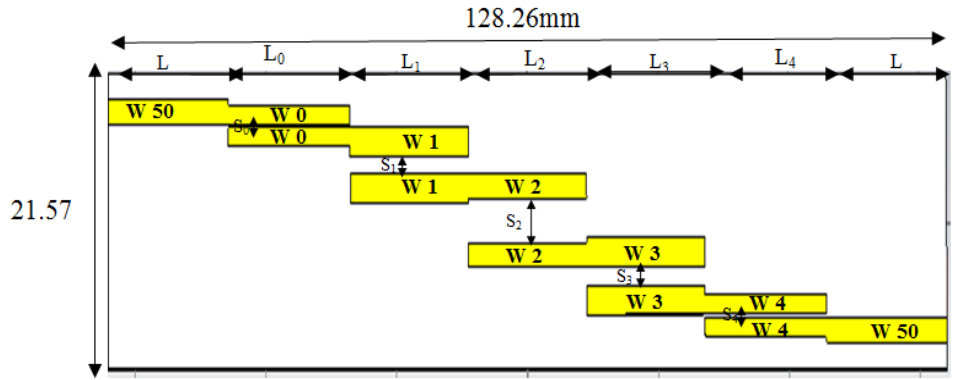

**Figure 5:** Parallel-coupled line layout using R03003.

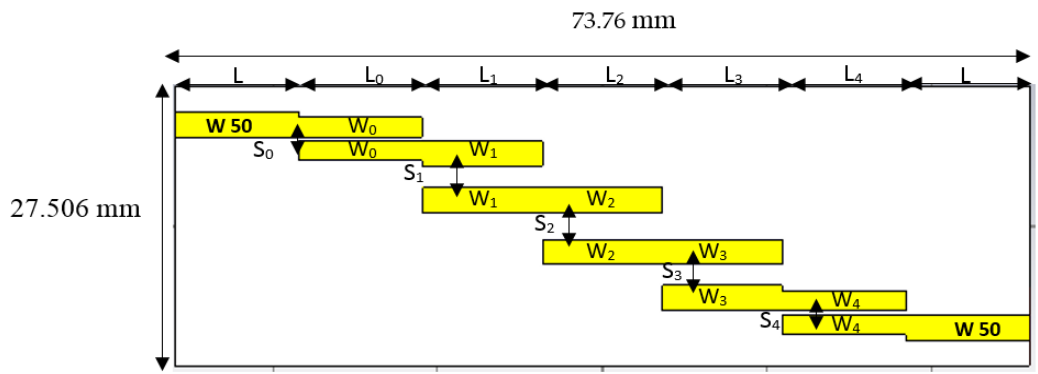

**Figure 6:** Parallel-coupled line layout using FR4.

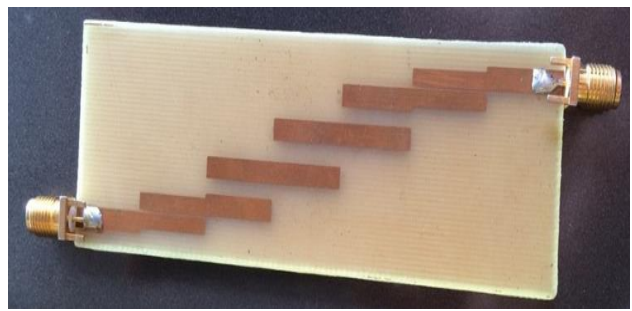

**Figure 7:** The fabricated PCL band-pass filter.

## 2.2 Multilayer Bandpass Filter Topology

The physical dimension of the filter starts with the determination of a multilayer stacking-up model that is used to construct the design. Three layers of the Printed Circuit Board (PCB) have been used to develop the design. Core material FR4 and epoxy material consists primarily of layers. The epoxy occupies the gap between core material and ground. Fig. 8 depicts the physical structure of the multilayer construction. Copper foil is laid down as PCB layer 3 followed by layer 2 (epoxy). The core material containing the top and bottom resonators is applied as upper layer 1. The circuit can be exposed itself to the air in a practical measurement. However, the part of air has been considered a vacuum in the simulation procedures, since the circuit has been simulated on the assumption that the circuit has been measured at a close boundary.

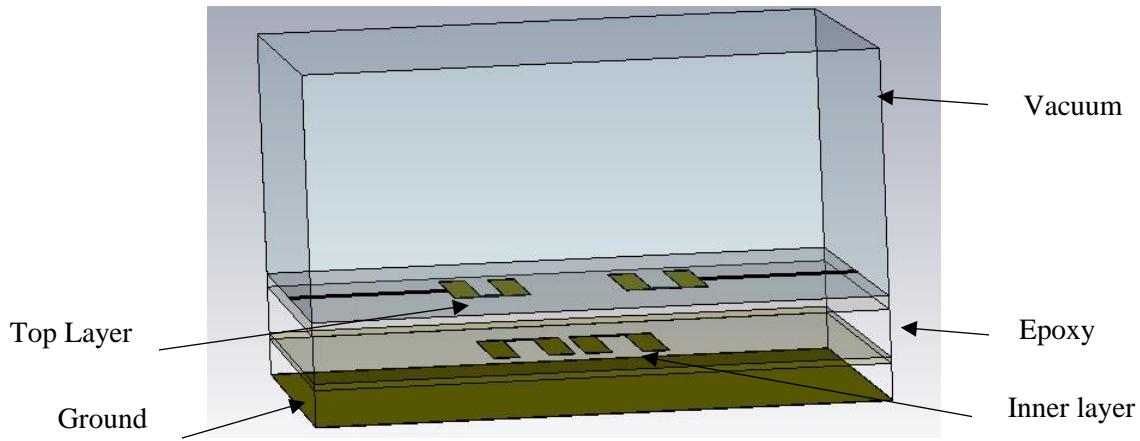

**Figure 8:** Construction of multilayer hairpin band-pass filter.

The filter design is derived simply from the design of the microstrip resonator. The adjacent resonator has been overlapped on a different layer, and strong couplings can be achieved between resonators. Fig. 8 shows the multilayer hairpin band-pass filter structure where resonator 1, 4 is placed on top of the core material, while resonators 2 and 3 have been placed on the bottom of the core material. A two-layer structure has been used to implement the resonators. Therefore, adjacent resonator lines are placed on different overlapping variations to the filter requirements obtained. The vacuum on the upper layer of a multilayer construction is considered to prevent or minimize the fringing effect as the signal propagates through the filter due to the boundary condition applied to each side of the filter dimension in CST. The multilayer structure has been combined with a thin epoxy layer, which acts as an adhesive material.

### *2.3 The Physical Layout of Multilayer Hairpin Bandpass Filter*

Several optimizations have been performed on the overall filter dimension to obtain the desired response, which meets the design specification. Figs. 9, 10, 11, and 12 show the filter's physical layout on the upper layer and bottom layer respectively after optimization. The overall filter size 38.66 mm x 31.41 using FR4 and the oversize using RO3003 is 47.13 mm x 37.42 mm.

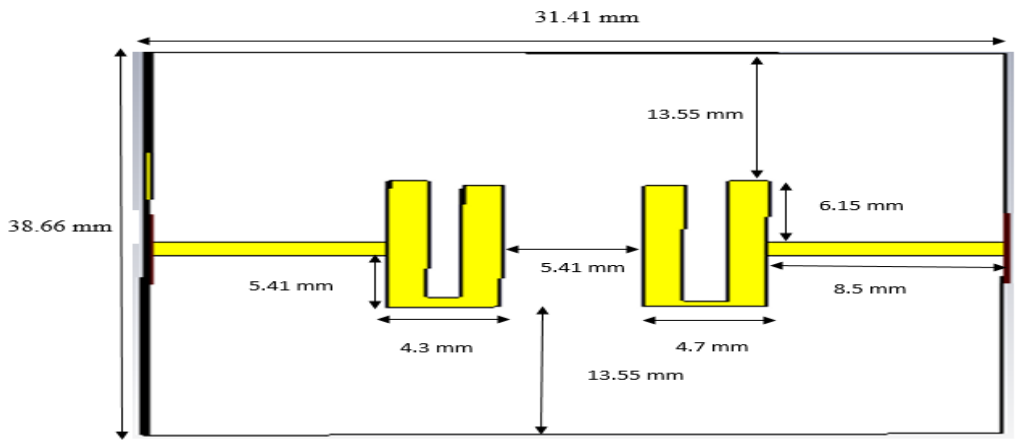

**Figure 9:** Physical layout on the top layer (FR4).

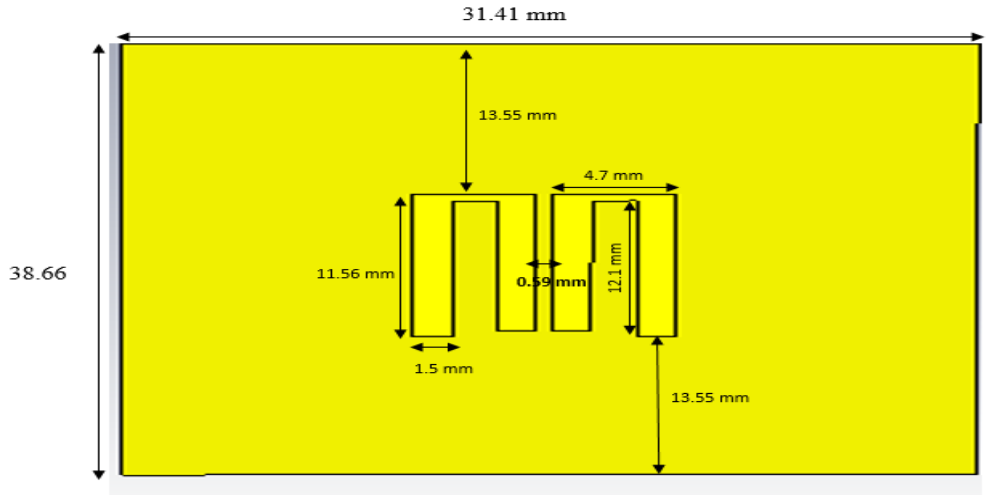

**Figure 10:** Physical layout on the inner layer (FR4).

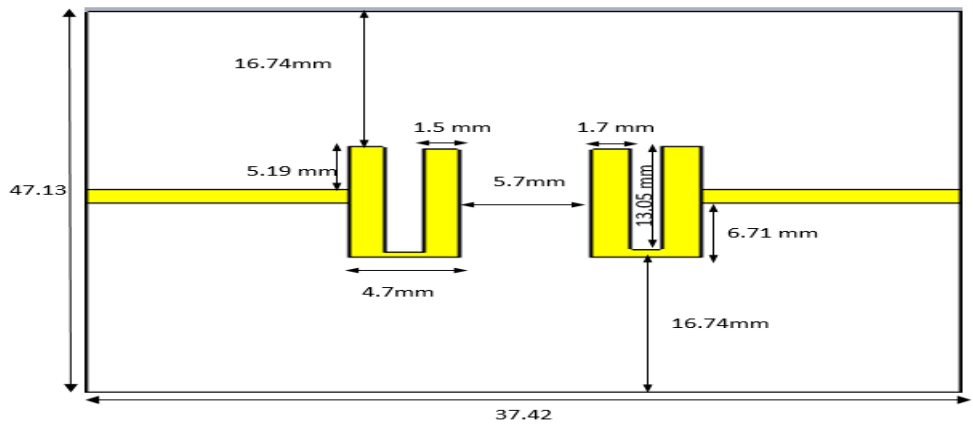

**Figure 11:** Physical layout on the top layer (RO3003).

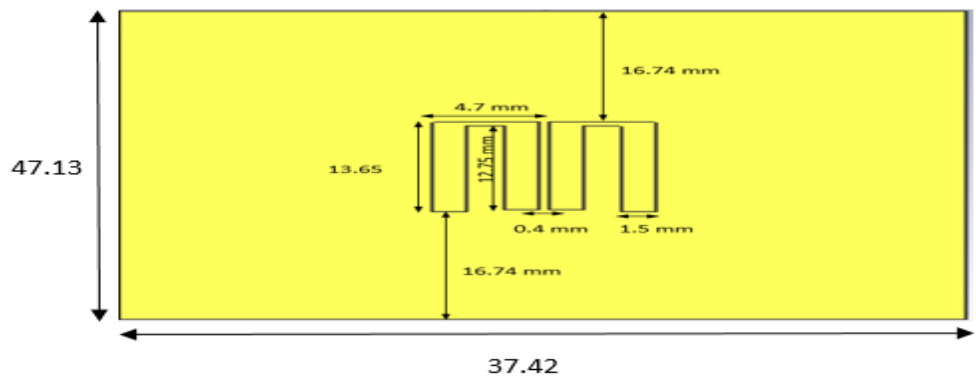

**Figure 12:** Physical layout on the inner layer (RO3003).

## 3. Results and Discussions
### *3.1 FR4 Based Design*

The proposed PCL filter has been designed and simulated using (ADS) software. The S-parameters ($S_{11}$ and $S_{21}$) simulation result of the PCL filter on frequencies from 2 GHz to 3 GHz is shown in Fig.13 using FR4. It can be observed from Fig.13 that ($S_{21}$) is less than -1.8 dB for the entire design frequency, while ($S_{11}$) is more than -30 dB.

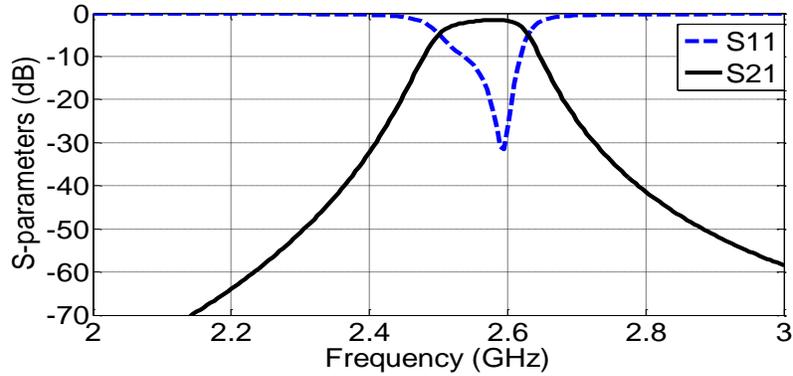

**Figure 13:** Simulated S-parameter (FR4).

In addition to this, the filter has been designed and simulated using commercially available (CST) software. It is required to optimize the filter physical dimensions to achieve the optimal required response. Fig.14 demonstrates a low insertion loss of - 4 dB and a reflection coefficient of better than -30 dB. The achieved BW of the proposed PCL filter is 210 MHz. The designed PCL filter has been fabricated on the FR4 substrate which has a permittivity of $\varepsilon_r$ = 4.3 and a physical thickness of 1.58 mm.

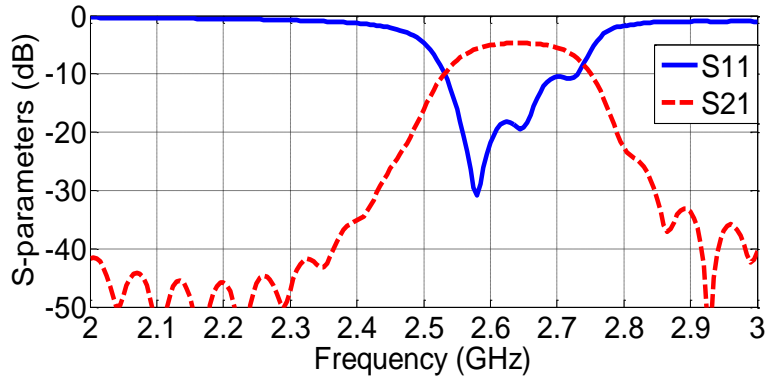

**Figure 14:** Simulated S-parameter.

Fig. 15 demonstrates the rendered PCL filter. The filter's scattering parameters have been measured via a Vector Network Analyser (VNA). The measured $S_{11}$ is compared with the EM simulation results, as depicted in Figure 15(a). The measured ($S_{11}$) is -39.24 dB at the center frequency(Fc) (2.64 GHz) and the measured ($S_{21}$) is -5.7 dB as shown in Fig. 15 (b). The simulated and measured S-parameter results agree well. However, there is some discrepancy in the measured and simulated results. These could be due to fabrication or experimental tolerances.

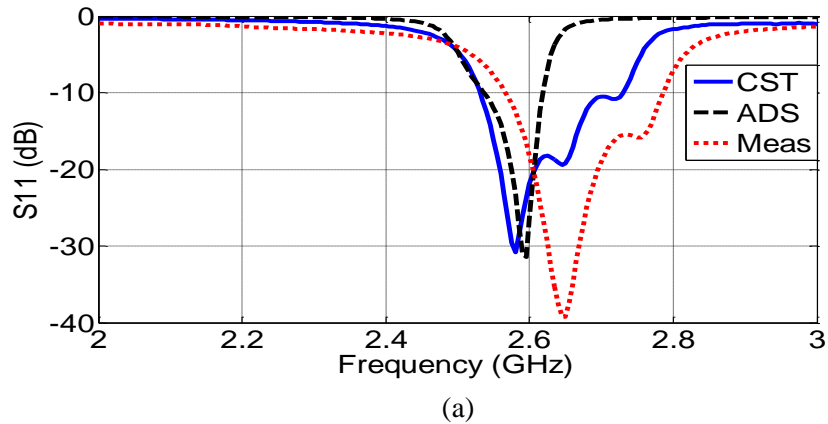

(a)

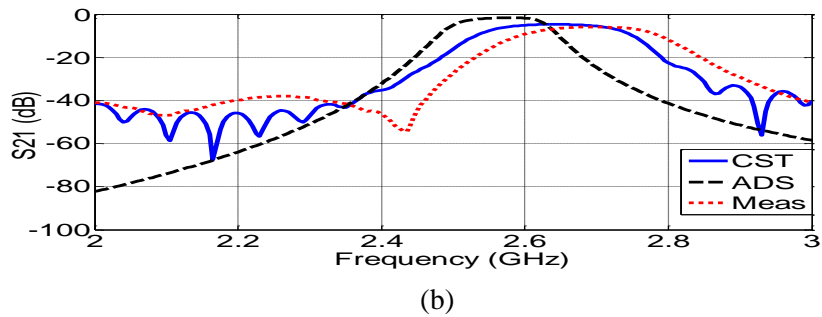

(b)

**Figure15:** Comparison of measured and simulated S-parameters of the PCL filter (**a**) $S_{11}$, (**b**) $S_{21}$ using FR4.

Fig. 16 shows the simulated response for a multilayer hairpin band-pass filter using the FR4 substrate of return loss ($S_{11}$) and insertion loss ($S_{21}$). Tab. 4 shows the performance of the filter for all parameters obtained from the simulated response. The ML filter has been designed and simulated with the aid of CST. The filter scattering parameters(S) have been calculated to achieve the target specification as tabled in Tab. 4. The simulated bandpass filter with a multilayer hairpin has a small insertion loss of -2.18 dB and a reflection coefficient of greater than -22.4203dB as shown in Fig. 16. More optimization of the dimensions is required to accomplish a better reflection coefficient and the same bandwidth. Other factors that may contribute to simulation errors are due to material loss, tangent loss of substrates and the adhesive epoxy used to join the filter layers. Nonetheless, the S-parameter results obtained are still reasonable concerning the design requirements and the ML filter structure discrepancies. Comparing with the PCL size which is 73.7615 mm x 27.50662 mm, the oversize has been reduced to 38.66 mm x 31.41 mm.

**Table 4:** Simulation results compared with filter design specifications

| Parameters | Simulation Values | Filter Specifications |
|---|---|---|
| Lower cut-off frequency, fU (GHz) | 2.5057 | 2.52 |
| Upper cut-off frequency, fL (GHz) | 2.6568 | 2.65 |
| ($S_{21}$) dB | -2.18 | > -3 |
| ($S_{11}$) dB | -22.4203 | < -20 |
| BW MHz | 151 | 130 |
| FC | 2.58 | 2.585 |

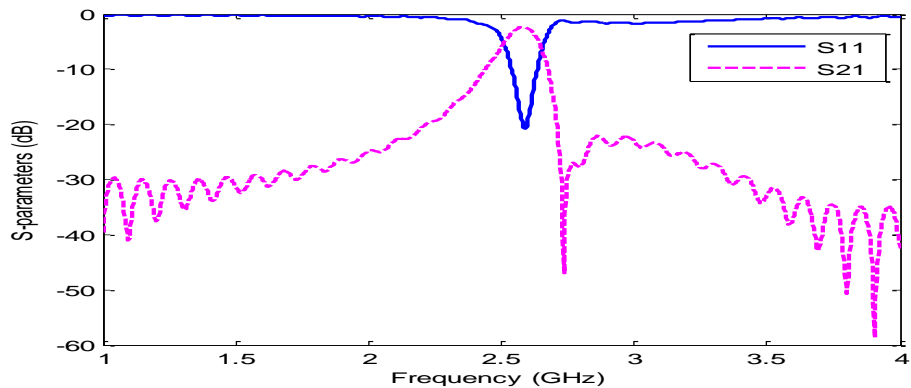

**Figure 16:** Simulated response of S-parameters (FR4).

*3.2 Rogers RO3003 Based Design*

The PCL filter has been simulated using ADS and CST. Fig. 17 and Fig. 18 show the simulated responses. The response in Fig.17 using ADS shows that $S_{11}$ is -24.625 dB and $S_{12}$ is -1.982dB. The result using CST in Fig. 18 shows that the $S_{11}$ has a low value equal to approximately -18.5 dB at the 2.58 GHz center frequency and $S_{21}$ is 1.79 dB. ML filter is simulated using CST and the simulated result in Fig. 19 shows a slight difference at the cut-off frequency that causes a reduction in the bandwidth. The response of $S_{11}$ is -18.993 dB and $S_{12}$ is -1.009dB.

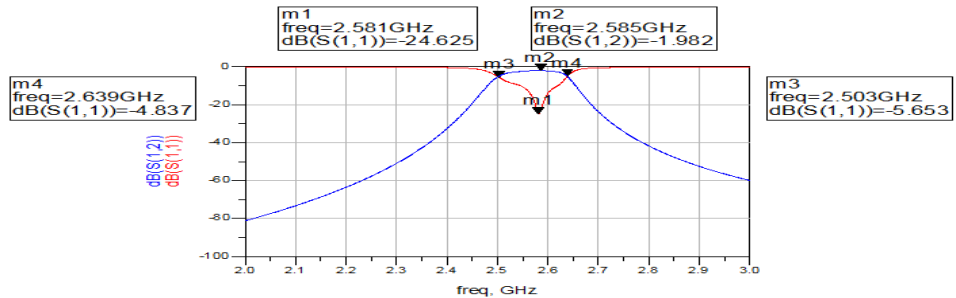

**Figure 17:** Simulated S-parameters of PCL filter using ADS (RO3003).

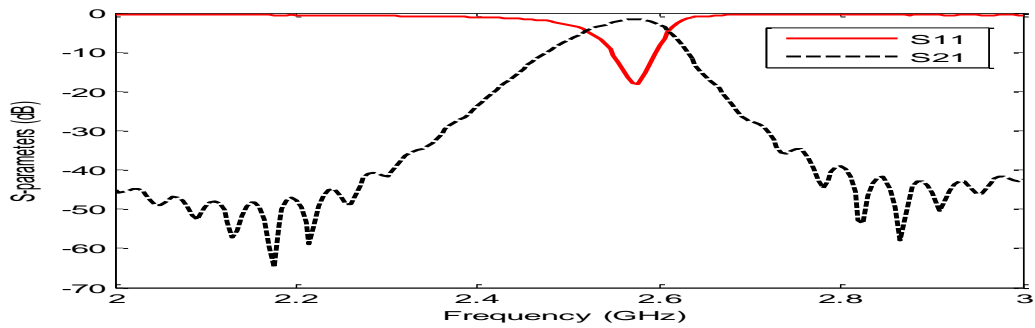

**Figure 18:** Simulated S-parameters of PCL filter using CST(RO3003).

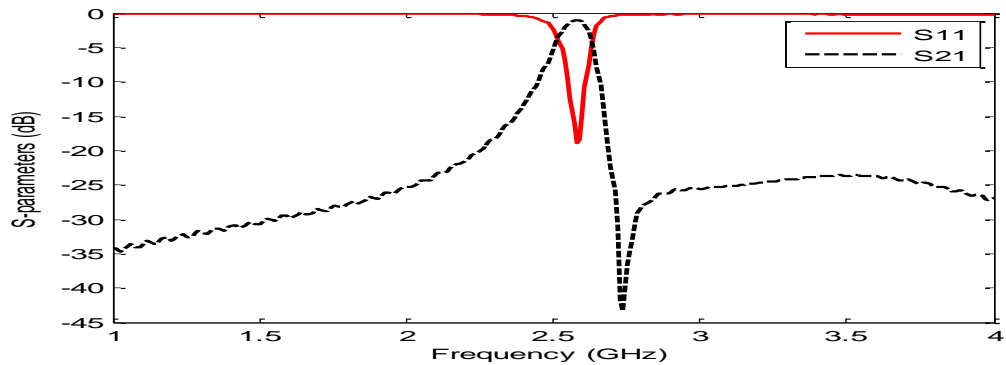

**Figure 19:** Simulated S-parameters of ML filter using CST(RO3003).

### *3. 3 Parametric Studies*

Fig. 20 and Fig. 21 show the PCL filter and ML Filter comparison of simulated S-parameters between FR4 and RO3003 substrate respectively. Because of the high dielectric constant of FR4, the size using FR4 is smaller than the design using Rogers RO3003. The wavelength is inversely proportional to the value of a dielectric constant. More compact size can be achieved, as the dielectric constant is small. Nonetheless, the overall parameters obtained from the simulation are well in line with the target specification set at the start of the design. The bandwidth using FR4 is larger, compared to RO3003 as shown in Figs. 20 and 21. However, some optimization is required to obtain a wider bandwidth as in the design specification. Nevertheless, the 2.58 GHz operating frequency can still be accomplished. Tab. 5 shows a comparison of simulation (FR4-RO3003) for PCL filter and Tab.6 shows a comparison of simulation (FR4-RO3003) for ML filter.

**Table 5:** Comparison of PCL filter simulation performance

| Properties | FR4 | RO3003 |
| --- | --- | --- |
| Lower cut-off frequency, $f_L$ (GHz) | 2.5297 | 2.5228 |
| Upper cut-off frequency, $f_U$ (GHz) | 2.7397 | 2.6111 |
| $S_{11}$ (dB) | -30.614 | -18.52 |
| $S_{21}$ (dB) | -5.651 | -1.79 |

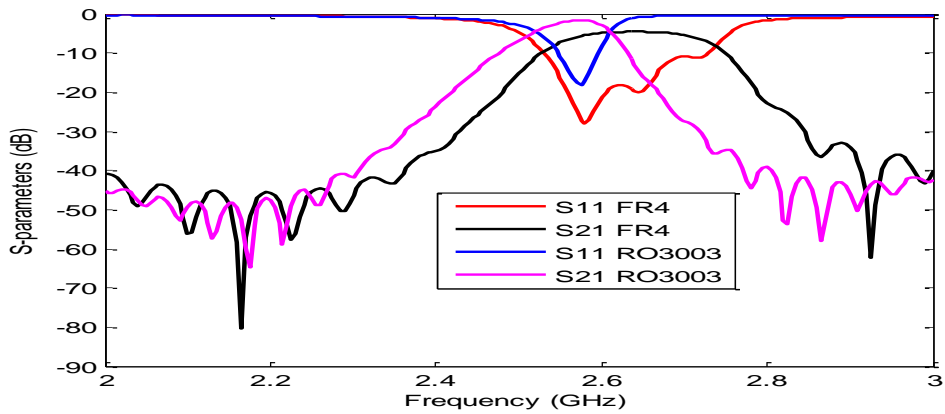

**Figure 20:** Comparison of simulated S-parameters PCL filter.

**Table 6:** Comparisons of ML filter performance

| Properties | FR4 | RO3003 |
| --- | --- | --- |
| Lower cut-off frequency, $f_L$ (GHz) | 2.5057 | 2.5172 |
| Upper cut-off frequency, $f_U$ (GHz) | 2.6568 | 2.6316 |
| FC (GHz) | 2.5851 | 2.5744 |
| $S_{21}$ (dB) | -2.18 | -1.009 |
| $S_{11}$ (dB) | -22.4203 | -18 |
| BW (MHz) | 151 (5.8449%) | 114.4 (5.148%) |
| Size (mm) | 38.6 × 31.41 | 47.13 × 37.42 |

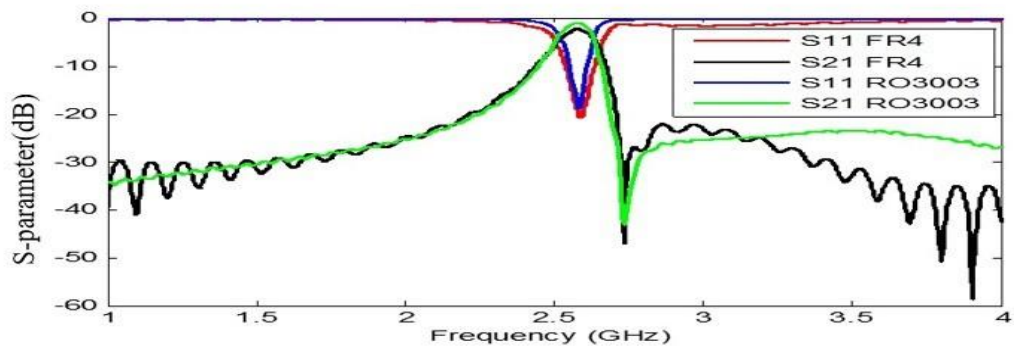

**Figure 21:** Comparison of simulated S-parameters ML filter.

**4. Conclusion and Future Development**

In this research, using microwave studio software (CST), PCL and ML bandpass filters have been proposed and analyzed. ML configuration has shown significant size reduction and compactness compared to PCL. The ML hairpin filters introduced a compact size because of their advantages compared to the PCL filter. The simulation results showed a good match with the required specifications. The analysis using FR4 which has different properties compared with RO3003 has proven the size of the filter can be smaller using a high dielectric constant substrate. For the filter to work well, high return loss and lower insertion loss are very critical factors to be taken into consideration. In terms of the performance of $S_{11}$ and $S_{21}$, the filter design using RO3003 is better compared to the filter design using FR4 substrate because of the low loss tangent of the substrate. It is due to the higher tangent loss of FR4, which results in a strong dissipation effect on the filter. In the future, this design can be improved by changing the core material with a smaller dielectric constant or smaller loss tangent. This change in core material can improve the response at the passband and extend the range of the filter. ML filter can be fabricated in the future and introduce transmission zeros at the stopband to improve the frequency selectivity of the filter.

**Funding Statement:** The authors received no specific funding for this study.

**Conflicts of Interest:** The authors declare that they have no conflicts of interest to report regarding the present study.